# Adaptive Adjustment of Relaxation Parameters for Algebraic Reconstruction Technique and its Possible Application to Sparsity Prior X-ray CT Reconstruction

Sajib Saha, Murat Tahtali, Andrew Lambert, Mark Pickering

University of New South Wales, Canberra, Australia

**Abstract:**

**Purpose:** In this paper, we systematically evaluate the performance of adaptive adjustment of the relaxation parameters of various iterative algorithms for X-ray CT reconstruction relying on sparsity priors. Sparsity prior has been found to be an efficient strategy in CT reconstruction where significantly fewer attenuation measurements are available. Sparsity prior CT reconstruction relies on iterative algorithms such as the algebraic reconstruction technique (ART) to produce a crude reconstruction based on which a sparse approximation is performed. Data driven adjustment of relaxation has been found to ensure better convergence than traditional relaxation for ART. In this paper, we study the performance of such data driven relaxation on a (CS) compressed sensing environment.

**Methods:** State-of-the-art algorithms are implemented and their performance analyzed in regard to conventional and data-driven relaxation. Experiments are performed both on simulated and real environments. For the simulated case, experiments are conducted with and without the presence of noise. Correlation coefficients, root mean square error, structural similarity index and perceptual dissimilarity metric were used for the quantitative comparisons of the results.

**Results:** Experiments reveal that data driven relaxation also ensures overall better quality reconstruction in a CS environment compared to traditional relaxation. However, when the data are corrupted by noise, inconsistencies emerge in the convergence unless a threshold is imposed on the maximum amendments.

**Conclusions:** Data driven relaxation seems a logical choice to more rapidly reach the solution. In a compressed sensing environment, especially when the data are corrupted by noise, a threshold to specify the maximum amendments needs to be specified. Experimentally, we have set the threshold as 20% of the previous value and thus have ensured more consistency in the convergence.

**Introduction**

The clinical use of CT has dramatically increased over the last two decades. In 2007, more than 68 million CT examinations were performed in the US alone, and an annual growth-rate of 10% has been observed over the last few years [1, 2]. As a radiation intensive modality, this increased use of CT has led to concerns on radiation induced genetic, cancerous and other diseases [2-4]. Since X-ray imaging is a quantum accumulation process, the signal-to-noise ratio (SNR) relates quadratically to the X-ray dose. "Given other conditions being identical, reducing the X-ray dose will degrade image quality"[5] – implies that the radiation dose cannot be reduced arbitrarily. The question of reconstructing acceptable CT images at a minimum radiation dose level is therefore a hot topic. Several approaches have shown promise in reducing CT radiation doses. One important option to reduce CT radiation doses includes the optimisation of CT parameters (e.g. tube voltage, tube current, pitch, and reconstructed section thickness) [6]. Other available options are the use of dual energy CT [7], the use of simultaneously active multiple sources [8] to decrease the number of acquisitions, and application of patient protection methods (e.g. automatic tube-current modulation) [6]. The optimization of image reconstruction algorithms provides an efficient dose reduction strategy by producing meaningful reconstructions from fewer number of X-ray attenuation measurements.

Aiming at a minimum number of projections for reconstruction, iterative approaches are nowadays gaining popularity over analytical ones. The recently proposed CS [9] theory puts iterative approaches a step forward for CT reconstruction. State-of-the-art CS based CT reconstruction algorithms first apply the algebraic reconstruction technique (ART) [10] to provide a crude reconstruction of the cross-section, which is then transferred to a compressed domain (gradient or wavelet) where sparse approximation is performed to ensure better reconstruction.

An adaptive version of the ART has been proposed in [11], where a data-driven adjustment of the relaxation parameters and amplitude constraints during the reconstruction procedure is advised. As can be seen from the experimental findings in

[11], that adaptive ART (AART) ensures better convergence thus leads to better quality reconstruction compared to plain ART. In a CS environment for X-ray CT, where both the ART and CS phases work in conjugation with each other, the performance evaluation of AART becomes incontestably important, which has been explored in this paper.

Here, we evaluate the performance of data driven relaxation proposed in [11] with traditional relaxation, in a sparsity prior reconstruction framework. X-ray data acquisitions are based on the simultaneous X-ray capture modality proposed f in our previous work [8]. Several state-of-the art sparsity prior CT reconstruction algorithms are implemented and their respective performances are analysed in regard to data-driven and traditional relaxation. Experiments are conducted both on simulated and real environment. For the simulated environment, both noisy and noiseless cases are considered.

**Algebraic Reconstruction Technique**

Algebraic reconstruction technique (ART), which is considered as an important class of iterative approaches, assumes that the cross section consists of an array of unknowns, and then sets up algebraic equations for the unknowns in terms of the measured projection data. In order to introduce the reader to ART, we will first show how we may construct a set of linear equations whose unknowns are elements of the object cross section. The iterative method for solving these equations will then be presented.

In Figure 1, an imaginary square grid is superimposed on the image $f(x, y)$, where $f_j$ denotes the constant value in the $j^{th}$ cell. $N$ such cells represent the unknowns $f$ of the problem to be solved. A finite set of $M$ projections ($P=\{p_1, p_2,..., p_M\}$ ) is obtained, where each projection $p_i$ is defined by

$$\sum_{j=1}^{N} w_{ij} f_j = p_i, \qquad i = 1, 2, \ldots, M \tag{1}$$

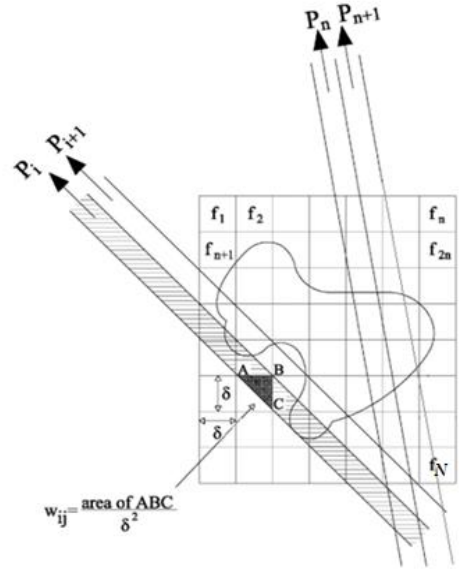

Figure 1. Image and projection representation for ART.

$w_{ij}$ is the weighting factor that represents the contribution of the $j^{th}$ cell to the $i^{th}$ ray integral as shown in Figure 1. $w$ is therefore a matrix of size $M \times N$. Thus, algebraic reconstruction algorithms try to find a solution to the system of equations in (1) relying on the following iterative scheme, which was discovered by Kaczmaz [12] long before its application to image reconstruction:

$$f^k = f^{k-1} + \beta \frac{(p_i - f^{k-1} w_i)}{w_i (w_i)^T} w_i \tag{2}$$

where $k$ is the iteration index, and $w_i = (w_{i1}, w_{i2}, \ldots, w_{iN})$. Here, $\beta$ is called 'relaxation parameter' that controls the convergence rate of the algorithm.

**Adaptive Adjustment of Relaxation Parameters**

To better understand the adaptive relaxation proposed in [11], lets rewrite equation (2) as

$$f^k = f^{k-1} + \lambda^{i,k}(p_i - f^{k-1}w_i), \text{ where } \lambda^{i,k} = \frac{w_i}{w_i(w_i)^T}\beta.$$

Here $i$ is the index of the current ray from 1 to $M$. In [11], the data-driven adjustment of $\lambda^{i,k}$ was defined as follows:

$$\lambda^{i,k} = \frac{x^{(k-1)}}{w_i x^{(k-1)}}\beta, \text{ where } x^{(k-1)} = [x_1^{(k-1)}, x_2^{(k-1)}, ..., x_N^{(k-1)}]^T, \text{ and } x_j^{(k-1)} = f_j^{(k-1)} w_{ij}.$$

As pointed out in [11], $w_{ij}$ only represents the geometry contribution of the *j*th pixel to the *i*th ray integral. The true contribution of the *j*th pixel to the *i*th ray integral is $x_j^{(k-1)}$. It is more reasonable to adjust the pixels that have a larger contribution $x_j^{(k-1)}$ to the *i*th ray integral with a larger adjustment step. Thus the proposed adaptive adjustment of relaxation parameters in [11] leads to not only to speedy convergence but also high-quality reconstruction, as shown in the next section. It is worth mentioning that generally $\beta$ is called the relaxation parameter, however in [11], the whole of $\lambda^{i,k}$ was termed as relaxation parameters.

**Sparsity Prior CT Reconstruction**

Sparsity prior CT reconstruction algorithms provide an efficient dose reduction strategy by producing meaningful reconstructions from fewer number of X-ray attenuation measurements. The underlying principle is to solving an underdetermined system by incorporating the sparsity prior as a constraint. This development - the problem of sparse prior recovery and/or compressed sensing (CS)- can in fact be traced back to earlier papers from the 90s such as [13], and later to the prominent papers by Donoho and Huo [14], and Donoho and Elad [15]. Following the discovery that sparsity could enable the exact solution of ill-posed problems under certain conditions [16, 17], there has been a tremendous growth on efficient application of sparsity constraints for solving ill-posed problems [18-20] in applied mathematics, computer science, and electrical engineering.

Although the mathematical principal of sparsity prior/CS is quite promising, its relevance in CT imaging relies on whether CT images are sparse or not. If an image is not sufficiently sparse, the CS algorithms will not be directly applicable to the problem. Fortunately, in CS theory, one can apply a sparsifying transform to increase the sparsity. The discrete gradient transform and wavelet transforms are frequently used to ensure the sparsity of the CT images.

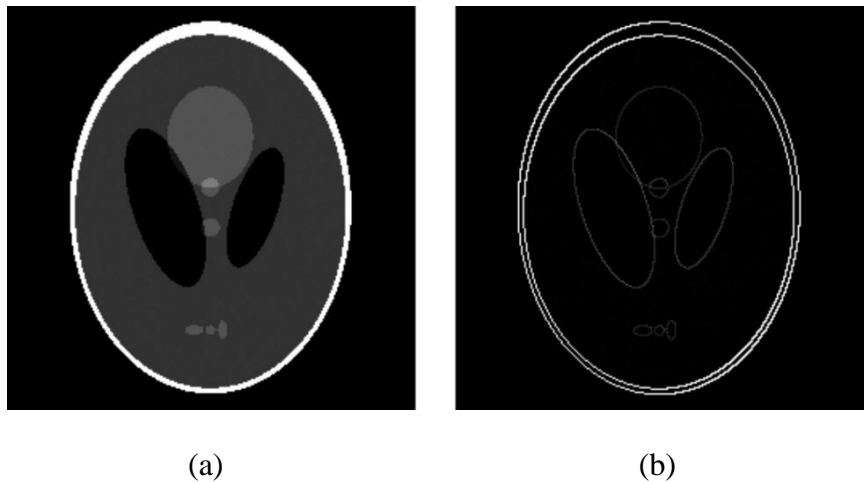

(a)          (b)

Figure 2. (a) Shepp-Logan Phantom. (b) The gradient counterpart of (a).

Hence the basic idea of compressed sensing based CT image reconstruction can be summarized as follows: instead of directly reconstructing a target image, the sparsified version of the image is reconstructed. After the sparsified image is reconstructed, an 'inverse' sparsifying transform is used to transform the sparsified image back to the target image. When the inverse-sparsifying transform is not explicitly available (such as for discrete gradient transform), an iterative procedure is used to perform the inverse-sparsifying transform during the image reconstruction process [21, 22].

Compressed sensing in CT prescribes solving the following constrained optimization problem:

$$\min ||\psi f||_1 \text{ subject to } ||Wf - p||_2^2 < \varepsilon. \tag{3}$$

Where $\psi$ is the known spasifying transform of $f$, which means that most entries of the vector $\psi f$ are zeros. $\varepsilon$ characterizes the raw data consistency. If the best reachable value of $\varepsilon$ is denoted by $\varepsilon_{opt}$, the value of $\varepsilon_{opt} > 0$ is not known before performing the iteration process [23]. Choosing $\varepsilon$ close to $\varepsilon_{opt}$ is recommended in [23] to guarantee a meaningful reconstruction.

Unconstrained formulation of equation (3) has been proposed by Song *et al.* and Yu *et al.* in [24] and [25], respectively:

$$\min ||\psi f||_1 + \mu ||Wf - p||_2^2. \tag{4}$$

For every value of $\varepsilon$ in equation (3) there exist a value of $\mu$ in equation (4) [23]. One benefit of using unconstrained formulation is that standard methods such as the conjugate gradient solver [26] can be used to solve the problem.

CS based methods in CT aim to improve the reconstruction quality and to decrease the image artefacts. A list of algorithms are already proposed in tomographic image reconstruction [21, 22, 27-33], where successful reconstruction depends on a proper initial guess, the regularization parameter, algorithms used for sparse approximation, the number of iterations, and so on. Sidky *et al.* [27, 28] proposed several implementations of a hybrid algorithm, the so called ADS-POCS framework, which treats the raw data fidelity and sparseness constraint separately in an alternating manner and produces meaningful reconstructions. Yu *et al.* [21] has shown that a local region of interest (ROI) can be exactly and stably reconstructed via the total variation (TV) minimization, provided that the object under consideration is essentially piecewise constant. Li *et al.* in [22] minimize the $l_1$-norm of the gradient image as the constraint factor for the iterative procedure. An adaptive version of the original PICCS algorithm, where the conventional CS objective function has been incorporated into the PICCS algorithm with a relative weighting, has been proposed in [30], that ensures higher image quality and reconstruction accuracy compared to other approaches. Lauzier *et al.* in [31] has demonstrated that a small ROI within a large object can be accurately and stably reconstructed provided that *a priori* information on electron density is known for a small region inside the ROI. An approach for solving the CT interior problem based on the high-order TV (HOT) minimization, assuming a piecewise polynomial ROI has been proposed in [32]. Cong *et al.* has shown in [33] through a numerical analysis framework that the accurate interior reconstruction can be achieved on an ROI from truncated differential projection data via the TV or HOT minimization, assuming a piecewise constant (polynomial) distribution within the ROI. However it is also important to ensure that the information about the image (or signal) is available and appropriately incorporated into the image reconstruction procedure, so that an image can be accurately reconstructed. In [34], a CS inspired rapid convergence of the iterative algorithm has been ensured through successful combination of several available/computed information. The algorithm starts with a good initial guess, uses contour information of the object, and relies on adaptive regularization to ensure rapid convergence. Its pseudo-code is given below:

Pseudo-code

Phase 1:

For $k = 1$ to *ART* do

$$f^0 = f^{ini}$$

For $i = 1$ to $M$ do

$$f^i = f^{i-1} + \lambda^{i,k}(p_i - f^{i-1}w_i)$$

Case 1:

$$\lambda^{i,k} = \frac{w_i}{w_i(w_i)^T}\beta$$

Case 2: Data driven relaxation

$$\lambda^{i,k} = \frac{x^{(i-1)}}{w_i x^{(i-1)}}\beta, \quad \text{where} \quad x^{(i-1)} = [x_1^{(i-1)}, x_2^{(i-1)}, ..., x_N^{(i-1)}]^T,$$

$$\text{and} \quad x_j^{(i-1)} = f_j^{(i-1)}w_{ij}.$$

End For ($i$)

Phase 2:

$$f_{m,n}^{temp} = T^{-1}(f^M)$$

For $P_{CG}=1$ to $P_{CG}$ do

Compute the search direction $d_{m,n}$:

$$d_{m,n} = \frac{4f_{m,n} - f_{m+1,n} - f_{m-1,n} - f_{m,n+1} - f_{m,n-1}}{\mu_{m,n}} +$$

$$\frac{f_{m,n} - f_{m+1,n}}{\mu_{m+1,n}} + \frac{f_{m,n} - f_{m-1,n}}{\mu_{m-1,n}} + \frac{f_{m,n} - f_{m,n+1}}{\mu_{m,n+1}} + \frac{f_{m,n} - f_{m,n-1}}{\mu_{m,n-1}}$$

where

$$\mu_{m,n} = \sqrt{\frac{(f_{m+1,n} - f_{m,n})^2 + (f_{m,n} - f_{m-1,n})^2 + (f_{m,n+1} - f_{m,n})^2 + (f_{m,n} - f_{m,n-1})^2}{2\nabla^2}}$$

$$\beta = \max(|f_{m,n}|) \div \max(|d_{m,n}|)$$

$$f_{m,n} = f_{m,n} - \alpha \times \beta \times d_{m,n}$$

$$\alpha = \alpha \times \alpha_s$$

End For ($P_{CG}$)

$$f^{ini} = T(f_{m,n})$$

End For ($k$)

Here, $T(I_{2d})$ is a function that transforms $I_{2d}$ to a 1-dimensional vector, similarly $T^{-1}(I_{1d})$ transforms a 1-dimensional vector $I_{1d}$ of $m \times n$ elements back to a 2-dimensional array of $m$ rows and $n$ columns.

For the experiment in this paper, $\beta$, $\alpha$, $\alpha_s$, and $\nabla$ are set respectively as 0.5, 0.005, 0.997 and 1, likewise in [21, 22, 30, 34]. Slice-by-slice reconstruction modality was considered.

**Review of the Simultaneous X-ray CT Acquisition Model**

Simultaneous X-ray acquisition model relies on the principle of lightfield imaging [35, 36], to also register the directional information of simultaneously X-rays. To replicate the concept of lightfield imaging in X-rays, the model uses an array of pinholes made of a material impermeable to X-rays. This array of pinholes, when placed in front of the sensor, allows the recording of directional information of the incoming X-rays along with their photon intensities, as shown in Figure 3(b).

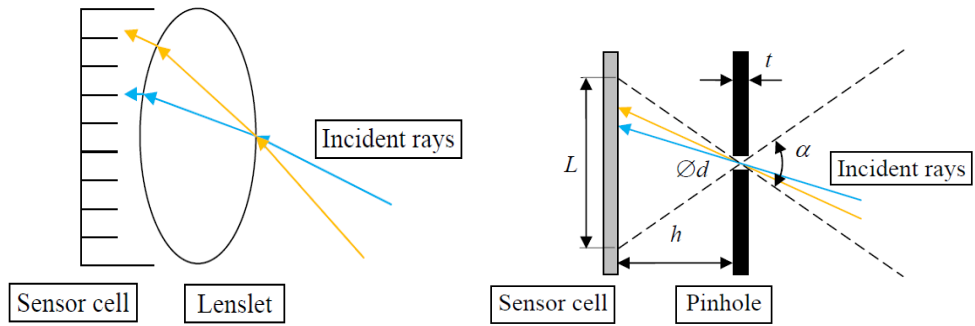

(a) (b)

Figure 3. (a) Each incident ray of different angle is captured by a different sensor element. (b) Pinhole acting as a lenslet - each incident ray of different angle is captured by a different sensor element.

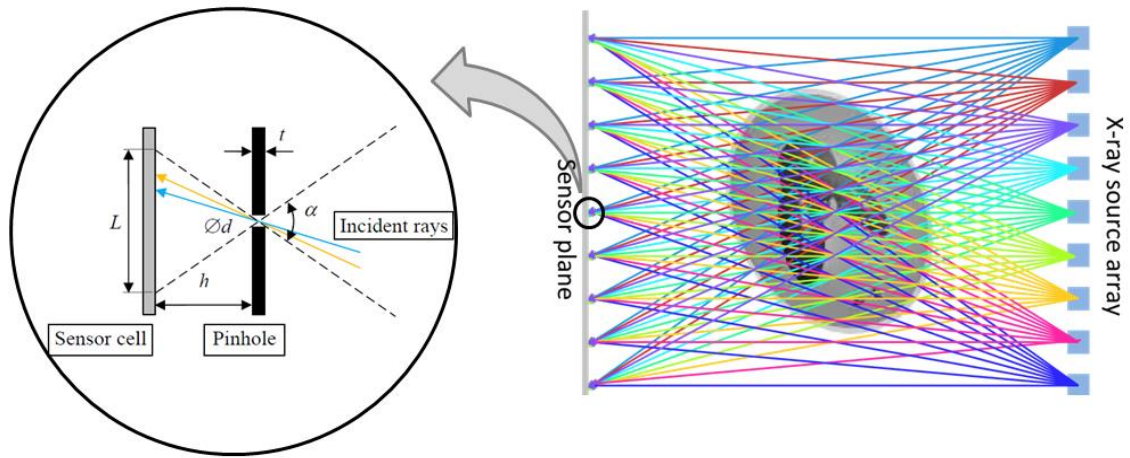

Figure 4. Simultaneously active multiple X-rays per projection.

Each angular projection consists of simultaneous X-ray exposures, as shown in Figure 4, at the same time multiple angular projections covering 180 degrees are considered. In this work, we consider 11 simultaneously active X-ray sources per angular orientation and eight angular projections at 0º, 22.5º, 45º, 67.5º, 90º, 112.5º, 135º and 157.5º are considered. The reason for this consideration is well explained in [37]. A trans-axial view of the considered setup is shown in Figure 5.

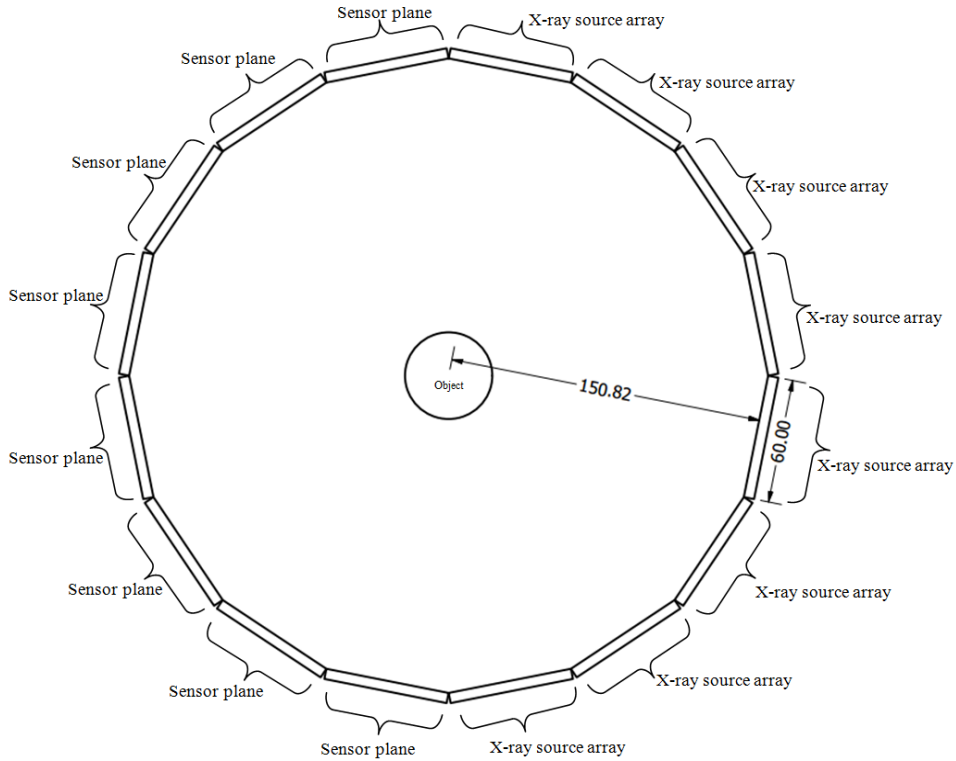

Figure 4. A trans-axial view of the simultaneous CT capture model–considers eight simultaneous angular projections.

**Experiments and results**

*Simulation Experiments*

A simulated environment was created in Matlab, where we considered eight angular projections at 0º, 22.5º, 45º, 67.5º, 90º, 112.5º, 135º, 157.5º and for each angular position we considered 11 simultaneously active X-ray sources. For the cross-section size of interest 50mm×50mm, the distance between the X-ray source array and the center of the object was set to 150 mm, so was the distance between the center of the object to the sensor plane. The sparsity prior algorithms proposed in [21, 22, 30, 34] are considered here as the state-of-the-art methods because of their generality in CT reconstruction, whereas others focus on CT reconstruction of specific organs. These algorithms are adapted to work in conjunction with the simultaneous X-ray acquisition model [8] and their respective performances are analysed in regard to traditional (case 1) and adaptive (case 2) relaxation parameters.

*Experiments to Evaluate the Overall Quality of Reconstruction*

Each projection datum along an X-ray through the sectional image is computed based on the known densities and intersection areas of the ray with the geometric shapes of the objects in the sectional image. After calculating the noise-free line integral $P_i$ as a direct projection operation, reconstructions were performed by the state-of-the art methods in comparison in regard to case 1 and case 2. Figure 6 shows the obtained reconstruction results after 10 iterations of ART (or such iterative algorithms) where each iteration of ART was followed by 5 compressed sensing iterations. The method proposed in [34] uses contour information of the object to eliminate all the pixels that lie outside the object's boundary. Hence, in order to have a fair comparison, we computed a binary mask having zero values outside the object's boundary and applied it on the reconstructed images produced by [21, 22, 30].

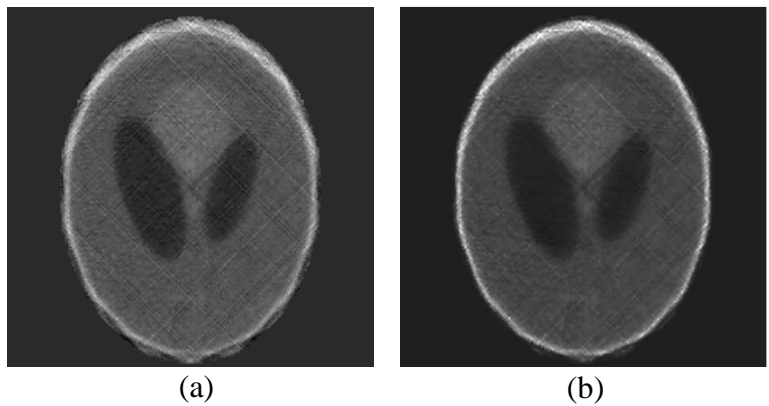

(a) (b)

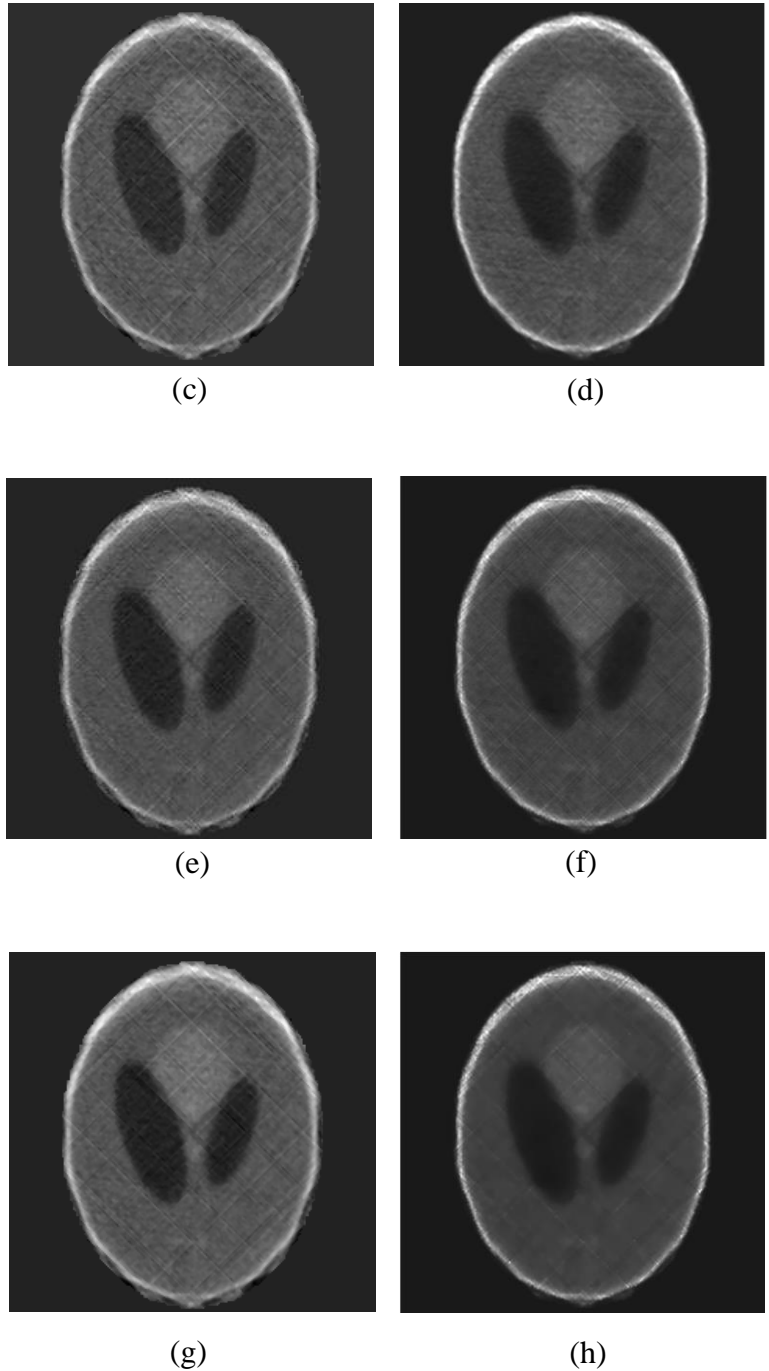

Figure 6: Reconstruction using (a) Yu et al. method (with traditional relaxation), (b) Yu et al. method (with adaptive relaxation), (c) Li et al. method (with traditional relaxation), (d) Li et al. method (with adaptive relaxation), (e) Lauzier et al. method (with traditional relaxation), (f) Lauzier et al. method (with adaptive relaxation), (g) Saha et al. method (with traditional relaxation), (h) Saha et al. method (with adaptive relaxation).

To compare the differences among various methods, intensity profiles of reconstructed images are drawn across the 128th column, from the 1st row to the 256th row. Figure 7 shows the intensity profiles of the reconstructed images by different algorithms in consideration and the corresponding profile of the original Shepp-Logan phantom.

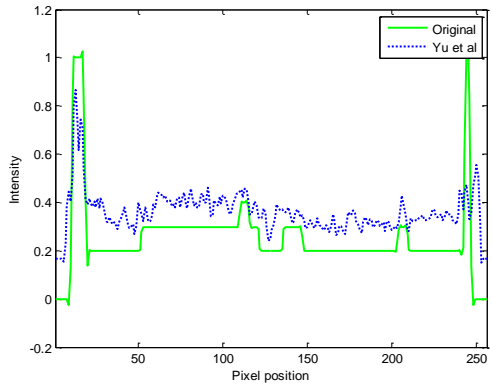

(a)

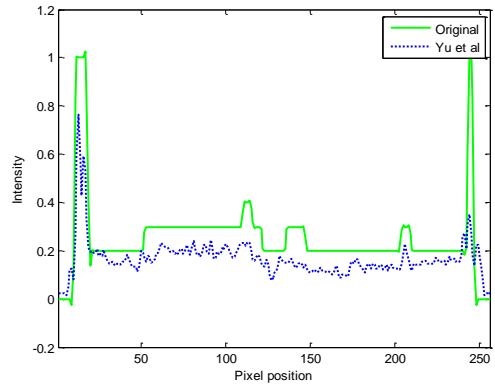

(b)

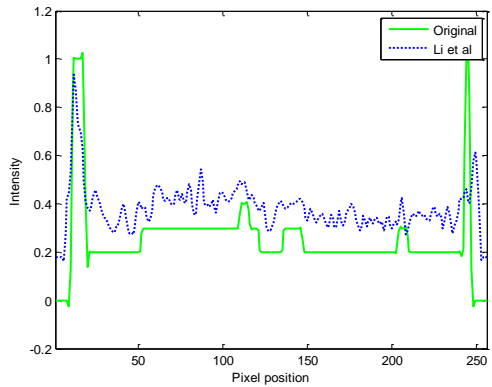

(c)

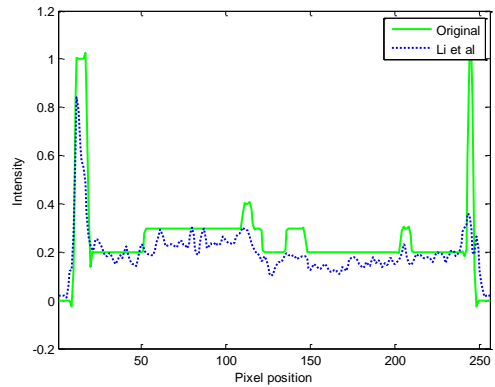

(d)

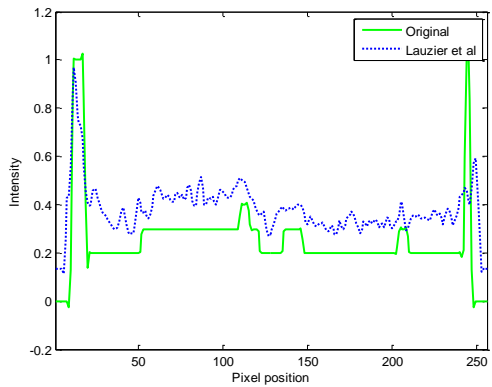

(e)

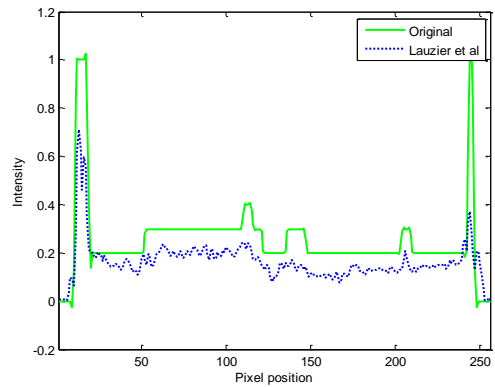

(f)

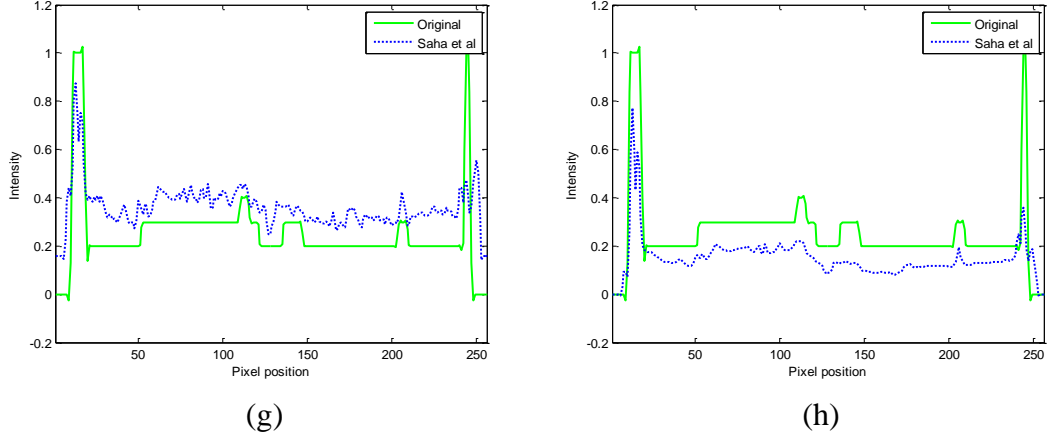

(g)   (h)

Figure 7: Pixel intensity profiles of the reconstructed images by (a) Yu et al. method (with traditional relaxation), (b) Yu et al. method (with adaptive relaxation), (c) Li et al. method (with traditional relaxation), (d) Li et al. method (with adaptive relaxation), (e) Lauzier et al. method (with traditional relaxation), (f) Lauzier et al. method (with adaptive relaxation), (g) Saha et al. method (with traditional relaxation), (h) Saha et al. method (with adaptive relaxation).

For the quantitative comparison of the methods in comparison, correlation coefficients ($coef(k)$), root mean square error ($rmse(k)$), SSIM [38] and PDM [39] were used. The mathematical expressions of $coef(k)$ and $rmse(k)$ are given below:

$$coef^{(k)} = \frac{\sum_i (t_i - \bar{t})(y_i^{(k)} - \bar{y}^{(k)})}{[\sum_i (t_i - \bar{t})^2 (y_i^{(k)} - \bar{y}^{(k)})^2]^{1/2}} \qquad (5)$$

$$rmse^{(k)} = [\frac{\sum_i (y_i^{(k)} - t_i)^2}{\text{Number of Elements in } t_i}]^{1/2} \qquad (6)$$

Here, $t_i(\bar{t})$ and $y_i(\bar{y}^{(k)})$ represents the pixel (average) value in the original and $k^{th}$ reconstructed images, respectively.

Figure 8 shows the intermediate reconstruction results for the different reconstruction methods in consideration.

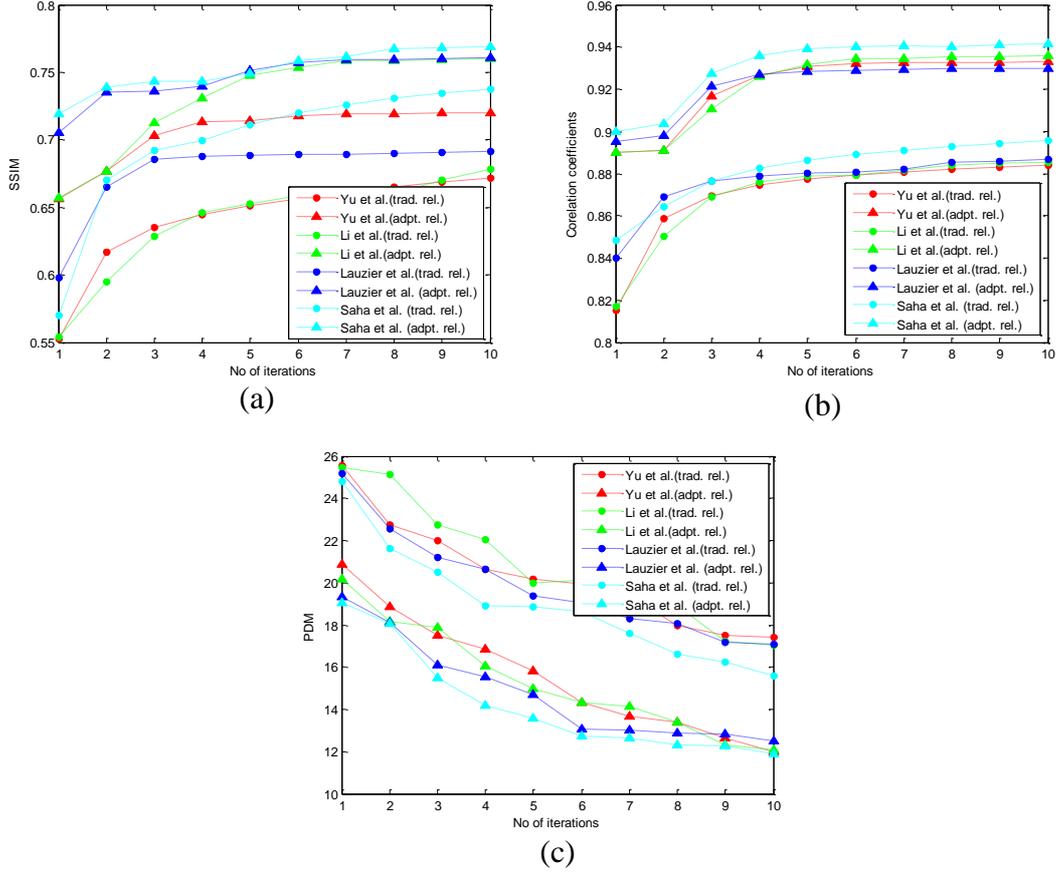

Figure 8: (a) SSIM, (b) Correlation coefficient, and (c) PDM between the original and reconstructed Shepp-Logan phantom.

The above experiment was performed without taking into account the effects of noise. In this part of the experiment, after calculating the noise-free line integral $P_i$ as a direct projection operation, the noisy measurement $b_i$ at each bin $i$ is generated following the statistical model of pre-logarithm projection data, as used in [40]:

$$b_i = \text{Poisson}(I_0 \exp(-P_i)) + \text{Normal}(0, \sigma_e^2). \tag{7}$$

Following Huang [40], the X-ray exposure level $I_0$ and the background electronic noise variance $\sigma_e^2$ are set to $9.0 \times 10^5$ and 10, respectively. Finally, the noisy measurement was obtained by performing the logarithm transform on $b_i$.

Figure 9 shows the reconstruction results with the presence of noise.

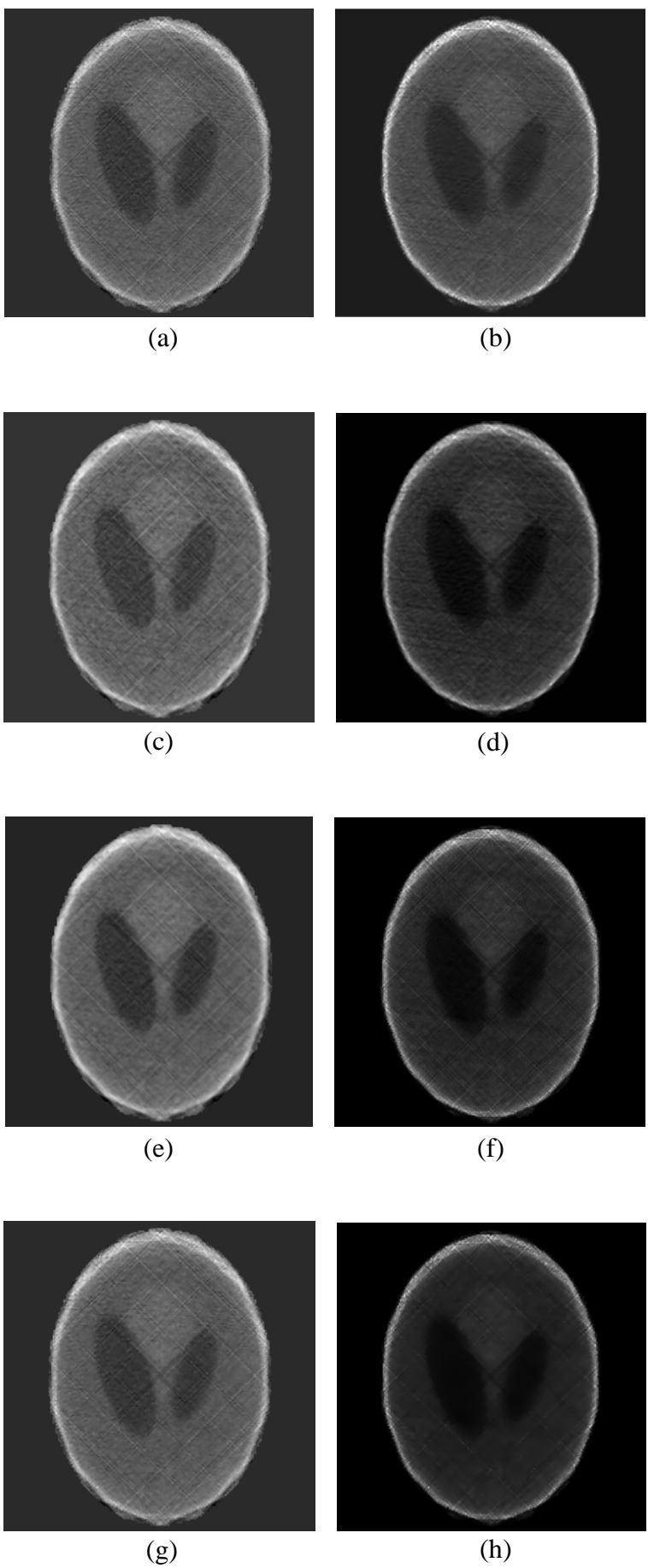

(a)

(b)

(c)

(d)

(e)

(f)

(g)

(h)

Figure 9: Reconstruction using (a) Yu et al. method (with traditional relaxation), (b) Yu et al. method (with adaptive relaxation), (c) Li et al. method (with traditional relaxation), (d) Li et al. method (with adaptive relaxation), (e) Lauzier et al. method (with traditional relaxation), (f) Lauzier et al. method (with adaptive relaxation), (g) Saha et al. method (with traditional relaxation), (h) Saha et al. method (with adaptive relaxation), with the presence of noise.

Figure 10 shows the intensity profiles and Figure 11 shows the intermediate reconstruction results.

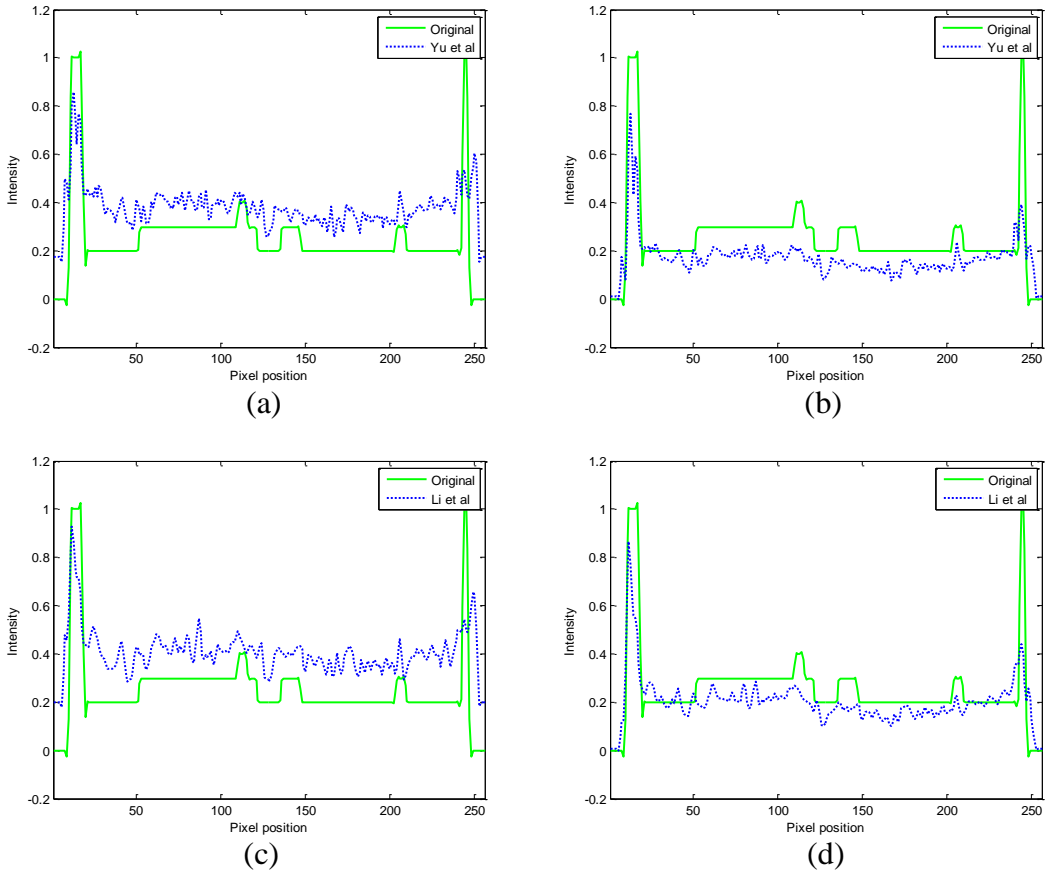

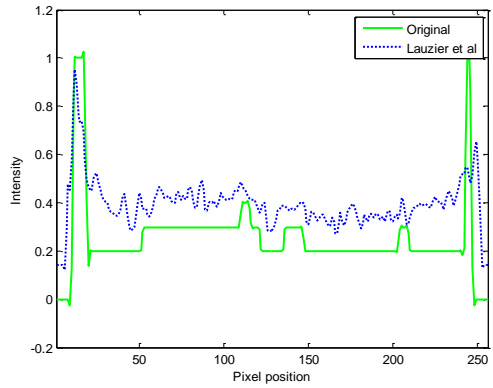 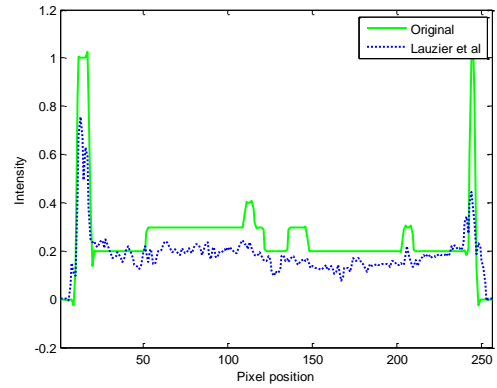

(e) (f)

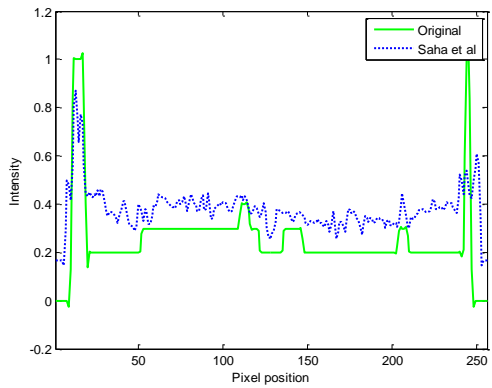 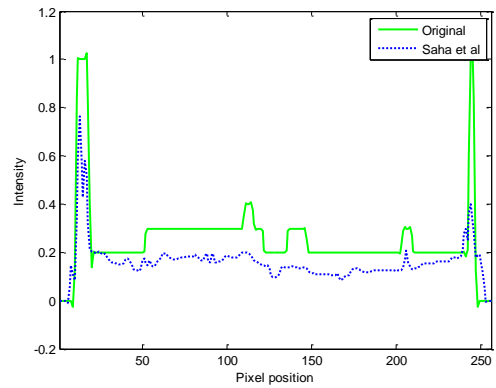

(g) (h)

Figure 10: Pixel intensity profiles of the reconstructed images by (a) Yu et al. method (with traditional relaxation), (b) Yu et al. method (with adaptive relaxation), (c) Li et al. method (with traditional relaxation), (d) Li et al. method (with adaptive relaxation), (e) Lauzier et al. method (with traditional relaxation), (f) Lauzier et al. method (with adaptive relaxation), (g) Saha et al. method (with traditional relaxation), (h) Saha et al. method (with adaptive relaxation), with the presence of noise.

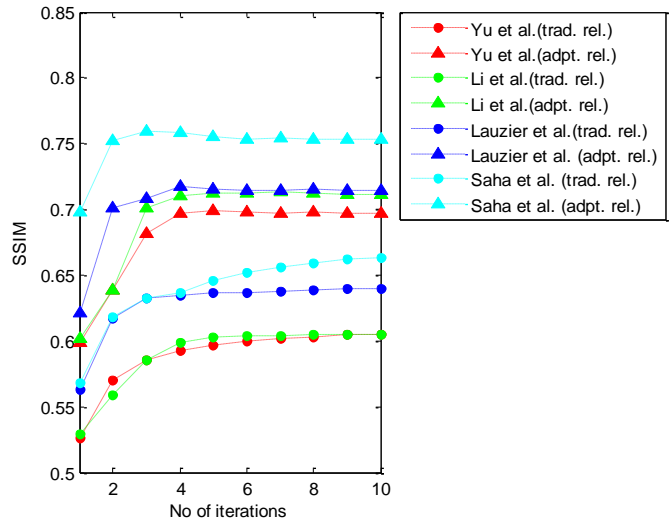

(a)

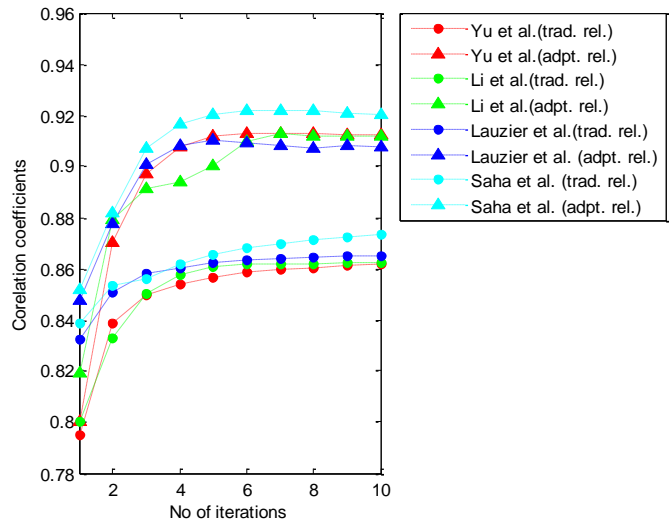

(b)

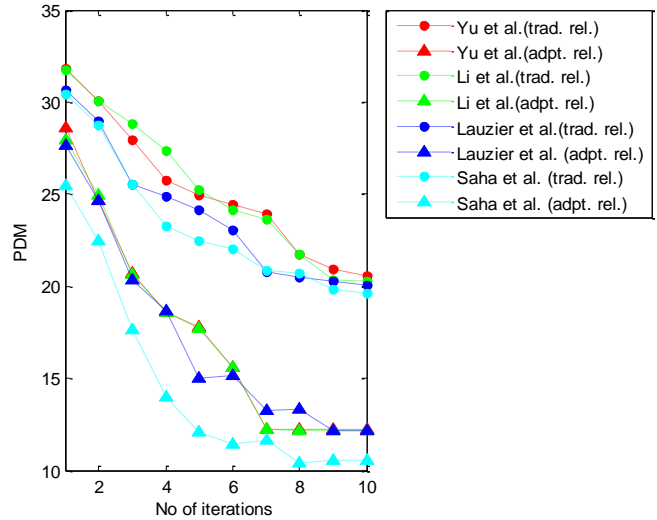

(c)

Figure 11: (a) SSIM, (b) Correlation coefficient, and (c) PDM between the original and reconstructed Shepp-Logan phantom, with the presence of noise.

The results imply that data driven relaxation ensures a better quality reconstruction than traditional relaxation. However, when the data are corrupted by noise, the results are not always convergent. As can be seen from figure 11, that with respect to the number of iterations, the SSIM index does not always increase or the PDM does not always decrease. For the initial few iterations, the results are convergent, however as soon as the iteration number increases, ups and downs in SSIM, PDM index rather than continuous improvement (for SSIM) or decrement (for PDM) are observed. Such inconsistencies in the convergence might affect defining the stopping criteria precisely based on achieved convergence after each iteration, which is a very common choice for iterative reconstruction algorithms. To our understanding, when a particular projection is severely affected by noise, the adjustments/amendments on the cross-sectional image by data-driven relaxation is more prone to be erroneous, and negatively affect convergence. Based on our findings, we have set a threshold $t$, such that for a particular image cell in consideration, data driven relaxation should not change the value more than *t*. Experimentally, we have set *t* as 20% of the previous value and have found more consistency in the convergence, as can be seen in Figure 12.

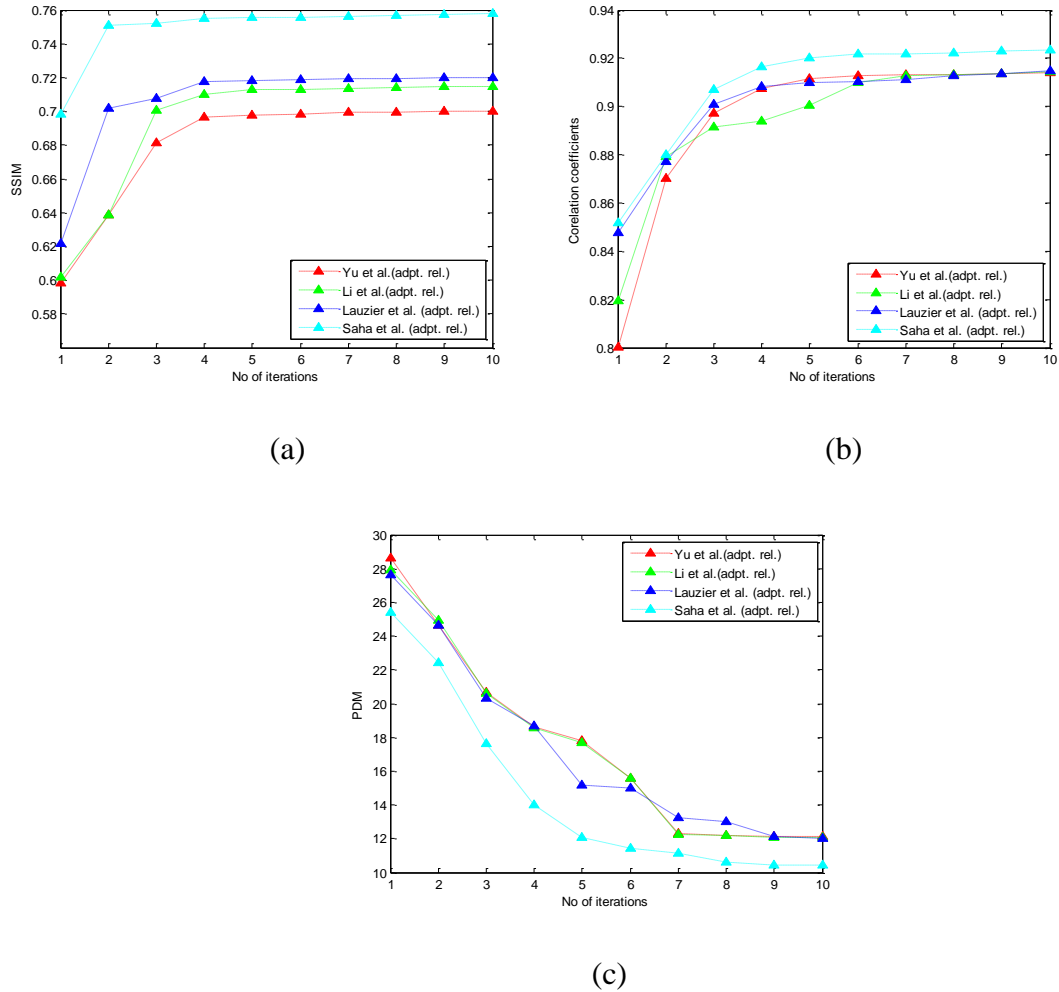

(a)

(b)

(c)

Figure 12: (a) SSIM, (b) Correlation coefficient, and (c) PDM between the original and reconstructed Shepp-Logan phantom, with the presence of noise. A threshold *t* as 20% of the previous value is imposed to specify the maximum amendments.

*Experiment on Real Data*

A physical 3D phantom, Figure 13, consisting of simple geometric shapes was used for the experiment. Eight angular projections were considered for the experiment, where for each angular orientation, 11 simultaneously active X-ray sources were considered.

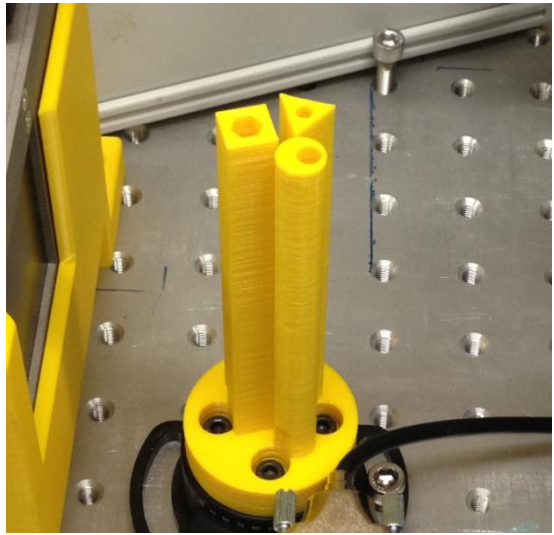

Figure 13. Photo of the physical 3D phantom used in the experimental.

Figure 14 shows a cross-section reconstructions by the methods in comparison on 256×256 grids.

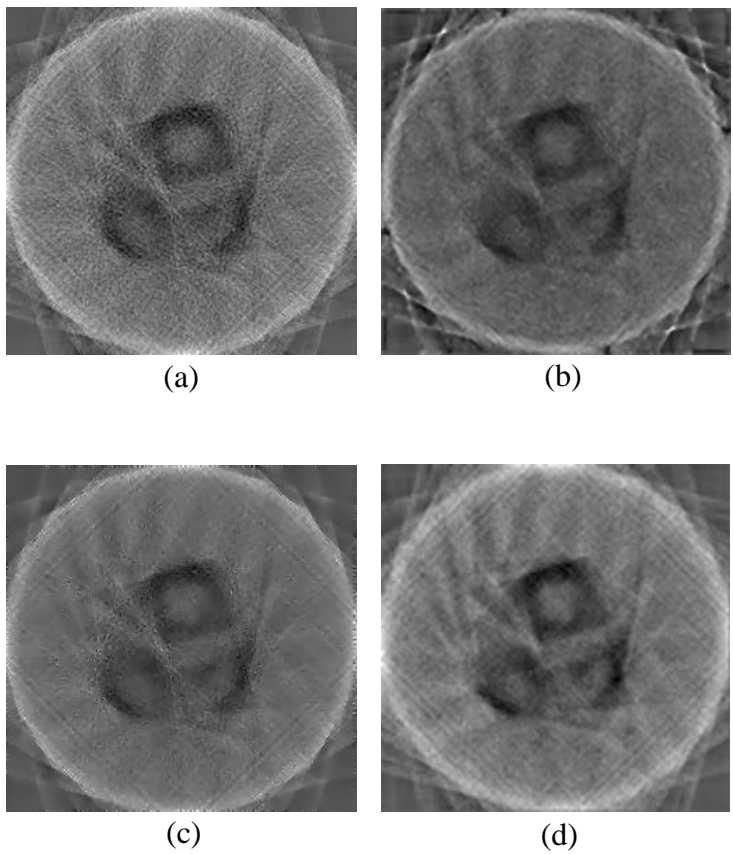

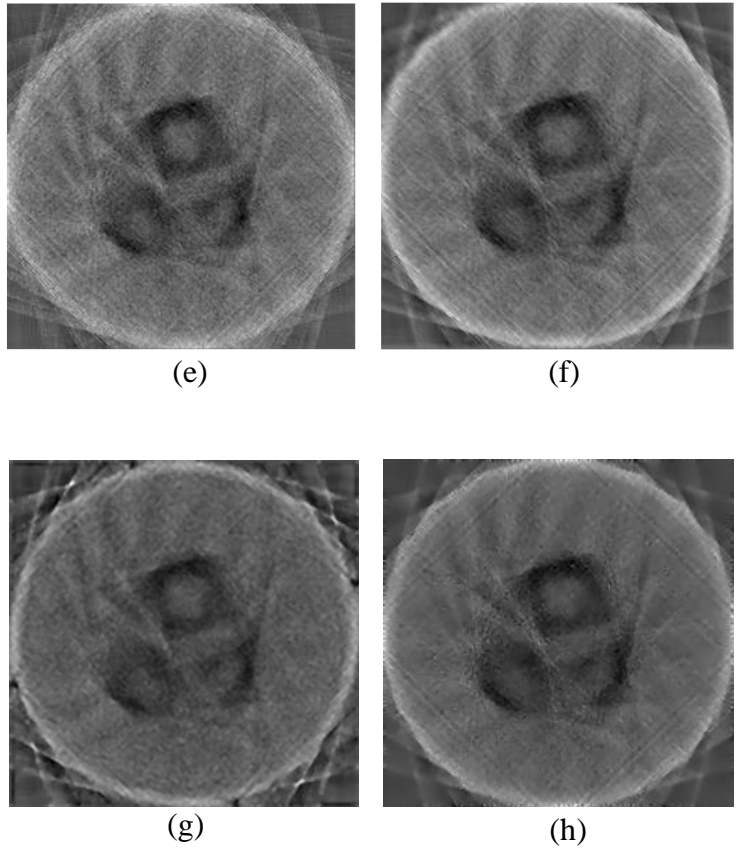

Figure 14: Reconstruction using (a) Yu et al. method (with traditional relaxation), (b) Yu et al. method (with adaptive relaxation), (c) Li et al. method (with traditional relaxation), (d) Li et al. method (with adaptive relaxation), (e) Lauzier et al. method (with traditional relaxation), (f) Lauzier et al. method (with adaptive relaxation), (g) Saha et al. method (with traditional relaxation), (h) Saha et al. method (with adaptive relaxation).

**Discussions and conclusion**

Compressed sensing or sparsity prior has gained significant attention in CT reconstruction because of its interesting capability to produce quality reconstructions with minimal number of attenuation measurements. It has been explored in the literature that CS-based iterative algorithms can yield images with quality compatible to that obtained with existing filtered back projection (FBP) and traditional iterative algorithms without sparsity prior, however with significantly less number of measurements. State-of-the-art CS based CT reconstruction algorithms first apply algebraic reconstruction technique (ART) [10] to ensure a crude reconstruction of the cross-section, which is then transferred to a compressed domain (gradient or wavelet) where sparse approximation is performed to ensure better reconstruction. More specifically, ART and CS work in succession repeatedly to improve reconstruction quality.

It has been explored in the literature that data driven relaxation, in comparison to traditional relaxation, ensures better convergence to ART. In this paper, we evaluate the performance of such data driven relaxation on a CS environment. X-ray acquisitions are based on the simultaneous CT capture modality proposed in our previous work[8, 41].

The state-of-the-art algorithms were implemented in Matlab from the provided pseudo-codes in the papers [20, 22, 30, 34]. All the experimental parameters were set as specified by the algorithm in consideration. Simulated and physical 3D phantoms were used for the experiment. For the simulated case, experiments were conducted both with and without the presence of noise. The results reveal that even in a CS environment, data driven regularization ensures better quality reconstruction than traditional relaxation. Although our naked eyes don't find very significant improvement in the reconstructed cross-section by data-driven relaxation, the improvements are evident through quantitative evaluations. One important finding is that when the data are corrupted by noise, the convergence of the sparsity prior reconstruction algorithms based on data driven relaxation becomes inconsistent. The proposed thresholding on the amount of amendments by the data driven relaxation seem to overcome such deficiencies at a certain extent.

It is evident that the reconstructions from real data are not as good as compared to the simulated data for all considered reconstruction methods. The most obvious reason is that for the simulation case, the forward problem (e.g. the projection data) was generated exactly following the principle that the X-ray path is exactly a straight and that there is no diffraction, nor weakening of the X-ray energy by the pinhole due to various interactions. Further work is require to take into account the omitted interactions to produce more realistic results for physical, whereas this wasn't the focus of the paper.